\documentclass[graybox]{svmult}

\usepackage{type1cm}        
\usepackage{makeidx}         
\usepackage{graphicx}        
\usepackage{multicol}        
\usepackage[bottom]{footmisc}
\usepackage{bbm}
\usepackage{newtxtext}       %
\usepackage[varvw]{newtxmath}       

\makeindex             

\begin{document}

\title*{Generative Adversarial Networks for Stain Normalisation in Histopathology}
\author{Jack Breen, Kieran Zucker, Katie Allen, Nishant Ravikumar, and Nicolas M. Orsi}
\institute{Jack Breen \at Centre for Computational Imaging and Simulation Technologies in Biomedicine (CISTIB), School of Computing, University of Leeds, UK, \email{scjjb@leeds.ac.uk}
\and Kieran Zucker \at Leeds Cancer Centre, St James’s University Hospital, Leeds, UK
\and Katie Allen \at Leeds Institute of Medical Research at St James's, School of Medicine, University of Leeds, UK
\and Nishant Ravikumar \at Centre for Computational Imaging and Simulation Technologies in Biomedicine (CISTIB), School of Computing, University of Leeds, UK
\and Nicolas M. Orsi \at Leeds Institute of Medical Research at St James's, School of Medicine, University of Leeds, UK
}
%
%
\maketitle

\abstract{
The rapid growth of digital pathology in recent years has provided an ideal opportunity for the development of artificial intelligence-based tools to improve the accuracy and efficiency of clinical diagnoses. One of the significant roadblocks to current research is the high level of visual variability across digital pathology images, causing models to generalise poorly to unseen data. Stain normalisation aims to standardise the visual profile of digital pathology images without changing the structural content of the images. In this chapter, we explore different techniques which have been used for stain normalisation in digital pathology, with a focus on approaches which utilise generative adversarial networks (GANs). Typically, GAN-based methods outperform non-generative approaches but at the cost of much greater computational requirements. However, it is not clear which method is best for stain normalisation in general, with different GAN and non-GAN approaches outperforming each other in different scenarios and according to different performance metrics. This is an ongoing field of study as researchers aim to identify a method which efficiently and effectively normalises pathology images to make AI models more robust and generalisable. 
}

\section{Histopathology}
\label{sec:1}
Histopathology is the microscopic evaluation of tissue for medical diagnosis. It is an essential part of the diagnostic pathway for many diseases, including autoimmune disorders, infections, and cancers. Tissue samples are taken either as biopsies or larger tissue resections, which are then fixed in formalin, embedded in paraffin, sectioned, and stained to create pathology slides. This process prevents tissue degradation and allows for long-term storage. Samples are occasionally flash-frozen, a faster process which allows pathologists to provide rapid information during surgery, but at the expense of increased cell damage and inferior staining quality. 
Samples are most commonly stained with haematoxylin and eosin (H\&E), which make cell nuclei appear blue and cytoplasm pink, with other cellular structures appearing varying shades of purple, pink, red, and blue. Immunohistochemistry (IHC) is another common staining technique which is used to identify specific antigens (proteins) in a tissue sample, which can aid in distinguishing between differential diagnoses and making prognostic predictions. 

Historically, pathologists analysed tissue samples using light microscopy; however, this is increasingly being replaced by a digital workflow where tissue slides are scanned at high resolution to generate whole slide images (WSIs) which can be visually assessed on a computer screen. Digitisation can drastically improve the efficiency of the diagnostic process \cite{Baidoshvili2018,Hanna2019} with minimal impact on diagnostic decisions \cite{Mukhopadhyay2018}, though the high start-up costs and technical requirements have slowed the rate of adoption. While the digital pathology workflow has primarily been developed for logistical and long-term financial reasons, it has also revolutionised diagnostic AI by creating large digital repositories of histopathology images. Models have been developed for a wide array of diagnostic and prognostic tasks 
\cite{Ahmed2022, Breen2023}, with AI researchers aiming to improve the accuracy and efficiency of the interpretation of pathology specimens. 
Diagnostic accuracy is essential to ensure that patients receive optimal treatment. Diagnostic efficiency is equally vital as there is a global shortage of pathologists, with some countries having access to only a small fraction of the number of pathologists available to others \cite{Wilson2018}, causing a diagnostic bottleneck. Diagnostic delays can have catastrophic consequences, with a four-week delay in cancer treatment being associated with a 10\% reduction in survival \cite{Hanna2020}. Demand for pathologists is expected to continue increasing due to the ongoing global population growth and ageing trends. 

Current AI models have limited clinical utility, with the United States Food and Drug Administration (FDA) having only approved one \emph{AI-enabled medical device} in digital pathology imaging. This tool classifies whether prostate biopsies contain malignant cells and indicates the most likely affected area \cite{Dasilva2021}. While this is a success story for digital pathology AI, the task of prostate biopsy malignancy classification has many enabling traits: it is a very common disease \cite{Sung2021}, biopsy slides contain orders of magnitude less tissue than resection slides, there are only two possible classes, and it is a relatively straightforward task for human pathologists, who achieve over 90\% accuracy \cite{Dasilva2021}. The high incidence rate of prostate cancer makes it possible to collect vast quantities of varied data, and the relatively small size of biopsy samples makes it possible to train a model with a huge number of samples, allowing for the development of a robust model. In scenarios where it is not possible to train a model with such varied data, model robustness is a critical barrier to clinical implementation. 

\begin{figure}[htbp]
\centering
\caption{Examples of visual variation caused by different scanners from the MIDOG 2021 Challenge training set \cite{Aubreville2023}, where each tissue sample was processed in the same laboratory following the same protocol, and then digitised with one of four available scanners. Image adapted from \cite{Breen2022}.}
\label{fig:midog}
\includegraphics[trim={0cm 0cm 0cm 0cm},width=\textwidth]{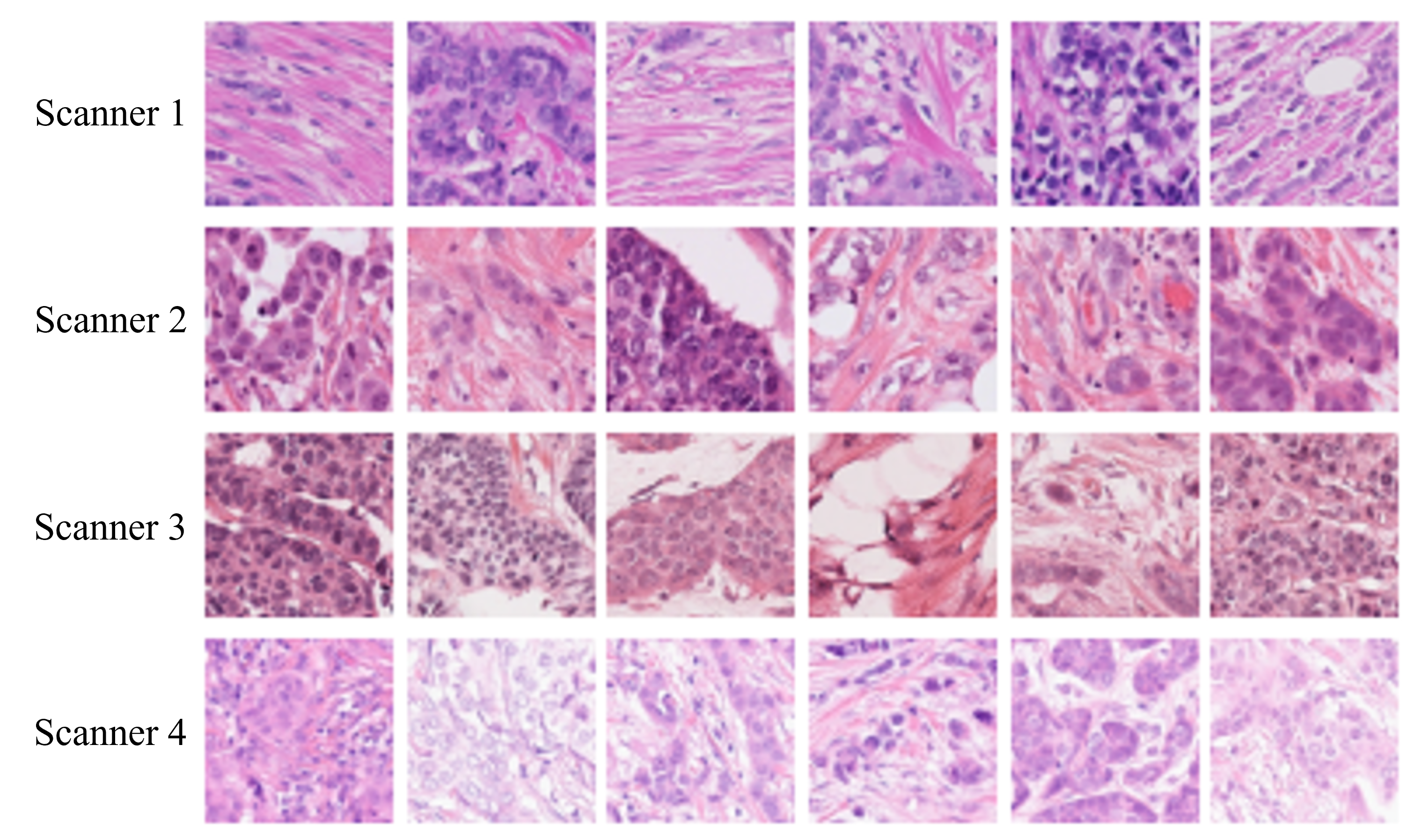}
\end{figure}

Digital pathology slides vary visually due to sample processing differences \cite{Lyon1994} (e.g. cut-up, fixation, and staining protocol), biological differences \cite{MIDOG2022} (e.g. tissue type, genetics), and digitisation differences \cite{Rajaganesan2021} (e.g. scanner, magnification, file formatting). Examples of the variation caused by different scanners are shown in Figure \ref{fig:midog}. Many factors will not vary significantly over short periods in a single pathology lab, meaning that single-centre data will typically be more homogeneous than multi-centre data, and hence that models trained on single-centre data will be less robust to variability. Such models are likely to generalise poorly to data from different data centres, and even to data from a single centre over time due to changes in tissue processing and digitisation. Stain normalisation, a form of style transfer, can improve model generalisability by reducing the variability of digital pathology data. 



\section{Style Transfer}

Style transfer methods aim to adjust the visual style of an image to match that of another image (or set of images) while retaining the original structure. 
Styles may be artificial, such as the painting style of a specific artist, or realistic, such as different seasonal colours in landscape photography. A toy example is converting ``horses to zebras" due to these animals sharing similar physiques but having distinct visual styles \cite{Zhu2017}. We may refer to these images as coming from different \emph{domains}, which in histopathology will often refer to different data centres which have distinct tissue processing and digitisation protocols. 

Style transfer may be supervised or unsupervised. Supervised training requires paired images showing the same content in different domains for training. 
Unsupervised training does not use paired images, so is often used in scenarios where it is impossible or impractical to collect perfectly paired images, such as the horse-to-zebra model and art-to-photograph models presented in the original CycleGAN paper \cite{Zhu2017}. 
It is not common to have paired histopathology data from different domains. Many domain differences are impossible to capture in paired histopathology data, such as the tissue staining procedure, which cannot be repeated for the same tissue sample. The variation of different scanners can be captured by repeatedly scanning the same sample, though this is rarely done as it increases costs without a direct benefit to patient care. Approximate paired data can be produced by differently processing consecutive tissue samples, though this is also rare. Supervised techniques can still be used when paired data is unavailable through the creative generation of artificial paired images, for example by taking the greyscale and full-colour versions of the same histopathology slide as a pair. This can be beneficial as supervised methods tend to be more computationally efficient and require less data than unsupervised methods.  



\subsection{Generative Models}
Many style transfer models are based on generative adversarial networks (GANs) \cite{Goodfellow2014}. These are two-part models containing a generator, which attempts to create new samples from a given distribution, and a discriminator, which attempts to distinguish real samples from generated ones. The standard GAN loss function may be expressed as:
\begin{equation} \label{eq:gan}
\mathcal{L}_{GAN}(G,D_G) = \mathbb{E}_{\mathbf{x}\sim p_{data}} [\log{D_G(\mathbf{x})}] + \mathbb{E}_{\mathbf{x}\sim p_{g}}[\log{(1-D_G(\mathbf{x}))}]\text{,}
\end{equation}
for generator $G$, corresponding discriminator $D_G$, real data distribution $p_{data}$, and generated data distribution $p_g$. This \emph{adversarial loss} represents the separation between real images and generated images, with the generator aiming to minimise the loss against its adversary, the discriminator, which aims to maximise it. 

\begin{figure}[htbp]
\centering
\caption{Simplified VAE-GAN architecture, with the VAE decoder used as the GAN generator. Figure adapted from \cite{Larsen2016}.}
\label{fig:vaegan}
\includegraphics[trim={0cm 1cm 0cm 0cm},width=\textwidth]{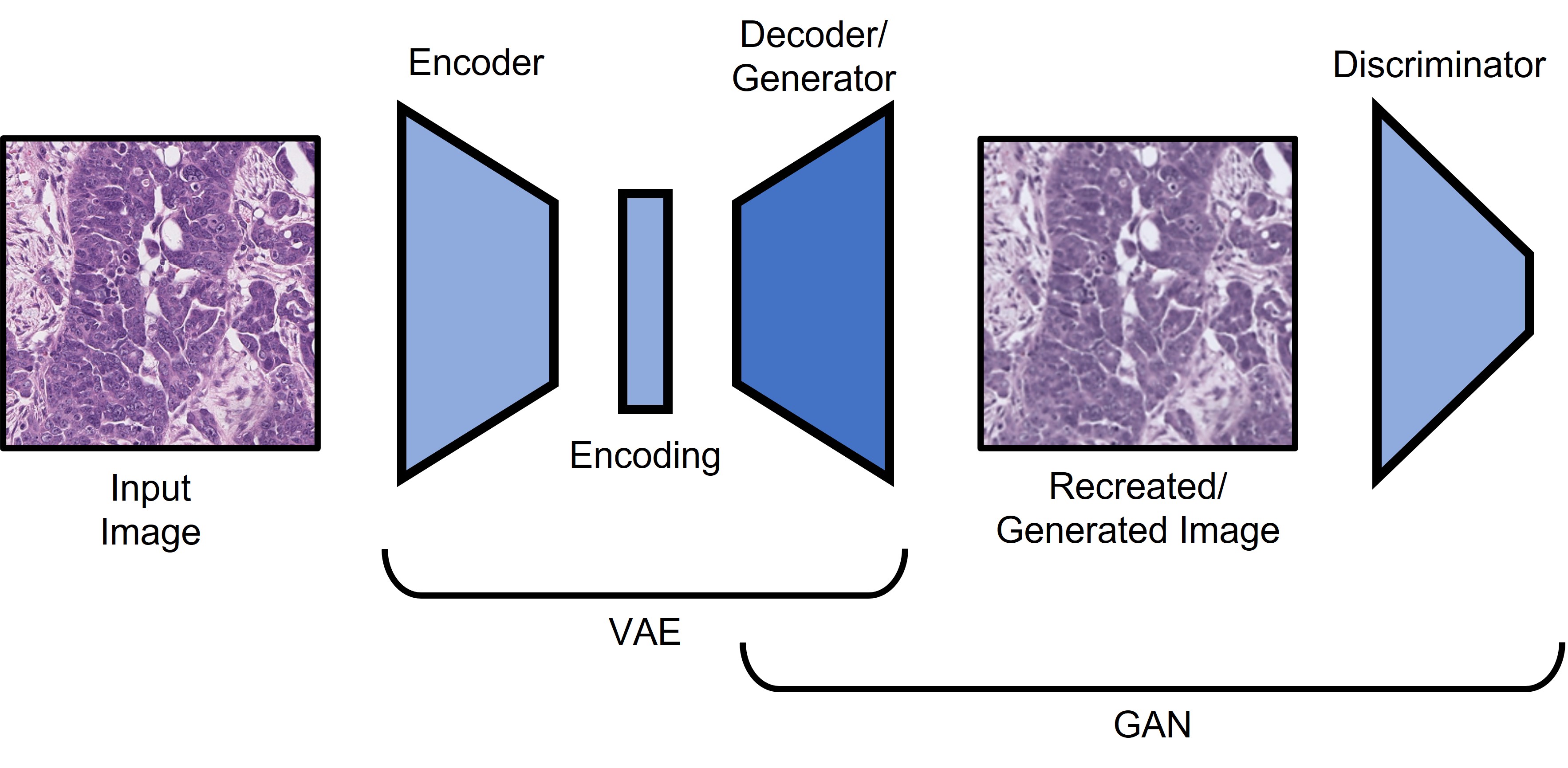}
\end{figure}

GANs are not the only generative models used in style transfer. Variational autoencoders (VAEs) \cite{Kingma2013} are encoder-decoder networks which impose a prior distribution onto the latent space. A latent loss is used during training to make the posterior latent distribution similar to the prior distribution. The standard VAE aims to exactly reproduce an input image, but it can be adjusted to produce style-transferred images, typically by combining it with a GAN to create a VAE-GAN, where the decoder of a VAE is used as the generator of a GAN, as shown in Figure \ref{fig:vaegan}.

Like VAEs, Gaussian Mixture Models (GMMs) assume an underlying distribution within the data, specifically a combination of multiple Gaussian distributions. In the context of style transfer, the typical approach is to iteratively adjust the colours of the input image using expectation maximisation to better fit the prior distribution. This approach has been applied to stain normalisation for many years \cite{Magee2009}, with newer approaches combining GMMs with deep learning \cite{Zanjani2018a}. 




\subsection{Similarity Metrics} \label{sec:metrics}
Generative style transfer methods can be evaluated in many ways. Some researchers focus on evaluating the effects of normalisation on a downstream AI task, while others try to directly evaluate the quality of synthetic images or the similarity between generated and real images. Stain normalisation can also be assessed qualitatively, for example by pathologists evaluating the quality of images and the severity of artifacts \cite{Michielli2022}.

Different quantitative similarity metrics are suitable in different scenarios, for example, if a model attempts to create an exact reconstruction of an input image then it may be appropriate to compare the original to the reconstruction using a holistic similarity measure such as Euclidean distance, L1 distance, or mean squared error. However, these distance metrics are less appropriate when comparing a stylised image to an original image, as the stylised image should not be an exact reconstruction of the input image. It may be more appropriate to use specific metrics such as the structural similarity index measure (SSIM), which focuses on shape but not colour. Maximising SSIM helps to find a model which retains the structural elements of an image, though this metric cannot quantify the change in style, which can be assessed qualitatively or through colour distribution comparisons (e.g. Kullback–Leibler divergence). 

Instead of directly comparing images, some metrics are based on the features generated by a convolutional neural network encoder. This includes the Fréchet inception distance (FID), which compares the distribution of encoded features from real and generated images, and the learned perceptual image patch similarity (LPIPS), which is explicitly trained to mimic human judgements. These metrics require careful training and are more computationally complex than non-trainable metrics, but they remain popular as they can offer more thorough assessments of similarity.

\subsection{Loss Functions}
Generative style transfer methods are often complex systems containing many different functions, each of which needs to be trained differently, leading to the use of multiple loss functions during the training of an individual network. These loss functions are diverse, although they typically include an adversarial loss and a reconstruction loss. The adversarial loss (such as the standard GAN loss in Equation \ref{eq:gan}) trains the model to produce realistic samples which fit within the distribution of the real images, and the reconstruction loss trains the model to retain certain information from the input image in the generated image. These are both important, as without the adversarial loss the model may not generate realistic images, and without the reconstruction loss the model may generate images which do not represent the input image. Models may also include a feature-preserving loss, which works the same as a reconstruction loss but is applied in an encoded feature space, and a conditional contrastive loss, which aims to cluster samples into groups with shared class labels in feature space. 

\subsubsection*{Reconstruction Losses}
Reconstruction losses are applied to retain information between an input image and a generated image. They are often based on the similarity metrics discussed in Section \ref{sec:metrics}, and can be as simple as taking a Euclidean distance between the input image $\mathbf{x}$ and generated image $G(\mathbf{x})$ \cite{Cho2017}: 
\begin{equation} \label{eq:euclidean}
\mathcal{L}_{r}(G) = \mathbb{E}_{\mathbf{x}\sim p_{data(x)}} [||\mathbf{x}-G(\mathbf{x})||_2]\text{.}
\end{equation}
Such a basic reconstruction loss can be prone to generating blurry areas, with the L1 loss mitigating this slightly but not solving the issue \cite{Isola2017}. These simple losses disincentivise any change between the original and generated image, so some researchers use more targeted metrics, such as SSIM, in their losses \cite{Liang2020}:
\begin{equation}\label{eq:ssim}
\mathcal{L}_{r}(G) = \mathbb{E}_{\mathbf{x}\sim p_{data(x)}} [1-SSIM(\mathbf{x},G(\mathbf{x}))]\text{.}
\end{equation}


\subsubsection*{Feature-Preserving Losses}
Feature-preserving losses work the same as reconstruction losses but are applied in an encoded feature space, rather than being applied directly to images. A simple Euclidean distance could still be applied as 
\begin{equation} \label{eq:fpeuclidean}
    \mathcal{L}_{fp}(G) = \mathbb{E}_{\mathbf{x}\sim p_{data(x)}} [||\mathscr{F}(\mathbf{x})-\mathscr{F}(G(\mathbf{x}))||_2]\text{,}
\end{equation}
where $\mathscr{F}(\mathbf{x})$ represents the features extracted from image $\mathbf{x}$. An example which has been used in this field is \cite{Cho2017}:
\begin{equation}\label{eq:fpkl}
\begin{split}
    \mathcal{L}_{fp}(G) &= \mathbb{KL}[\mathscr{F}(\mathbf{x}))||\mathscr{F}(G(\mathbf{x})))] \\ &= \mathbb{E}_{\mathbf{x}\sim p_{data(x)}} [\log(\mathscr{F}(\mathbf{x}))-\log(\mathscr{F}(G(\mathbf{x})))]\text{,}
\end{split}
\end{equation}
where $\mathbb{KL}$ is the Kullback–Leibler divergence. A similar loss is typically used in VAEs to impose the prior distribution onto the feature space \cite{Kingma2013}.

\subsubsection*{Conditional Losses}
Conditional GANs (cGANs) \cite{Mirza2014} leverage extra information, with the model depending on certain class labels associated with the images. This can be implemented in different ways, with the original cGAN approach \cite{Mirza2014} adjusting the standard GAN loss to give:
\begin{equation} \label{eq:cgan}
\mathcal{L}_{cGAN}(G,D_G) = \mathbb{E}_{\mathbf{x}\sim p_{data}} [\log{D_G(\mathbf{x}|\mathbf{y})}] + \mathbb{E}_{\mathbf{x}\sim p_{g}}[\log{(1-D_G(\mathbf{x}|\mathbf{y}))}]\text{,}
\end{equation}
where $\mathbf{y}$ are the class labels. Conditional losses may instead be applied in the feature space, where a conditional contrastive loss aims to create a feature space in which samples are tightly clustered by their input label according to a chosen distance/similarity metric, such as the Euclidean distance or cosine similarity. When using cosine similarity, the loss can be formulated as \cite{Kang2020}:
\begin{equation} \label{eq:ccl}
    \mathcal{L}_{ccl}(\mathbf{x}_i,y_i;t) = -\log\left(\frac{\exp(l(\mathbf{x}_i)^\top e(y_i)/t)+\sum_{k=1}^{m}\mathbbm{1}_{y_k=y_i}\cdot\exp(l(\mathbf{x}_i)^\top l(\mathbf{x}_k)/t)}{\exp(l(\mathbf{x}_i)^\top e(y_i)/t)+\sum_{k=1}^{m}\mathbbm{1}_{k\neq i}\cdot\exp(l(\mathbf{x}_i)^\top l(\mathbf{x}_k)/t)}\right)\text{,}
\end{equation}
for class embedding function $e$, input images $\mathbf{x}_i$ with corresponding class labels $y_i$, indicator function $\mathbbm{1}$, and scalar $t$.

\subsection{Multi-Generator Methods}
\begin{figure}[htbp]
\centering
\caption{Unsupervised image-to-image translation methods for images $x$,$y$ from domains $\mathcal{X}$,$\mathcal{Y}$, with encoders $E$, generators $G$, and latent spaces $\mathcal{Z}$. (a) CycleGAN \cite{Zhu2017} uses domain-specific latent spaces, (b) UNIT \cite{Liu2017} uses a shared latent space, (c) MUNIT \cite{Huang2018} and DRIT \cite{Lee2018} 
decompose the latent space into domain-specific attribute (style) spaces and a single shared content (structure) latent space. Diagram adapted from \cite{Lee2020}.}
\label{fig:imagetoimage}
\includegraphics[trim={0cm 0cm 0cm 0cm},width=\textwidth]{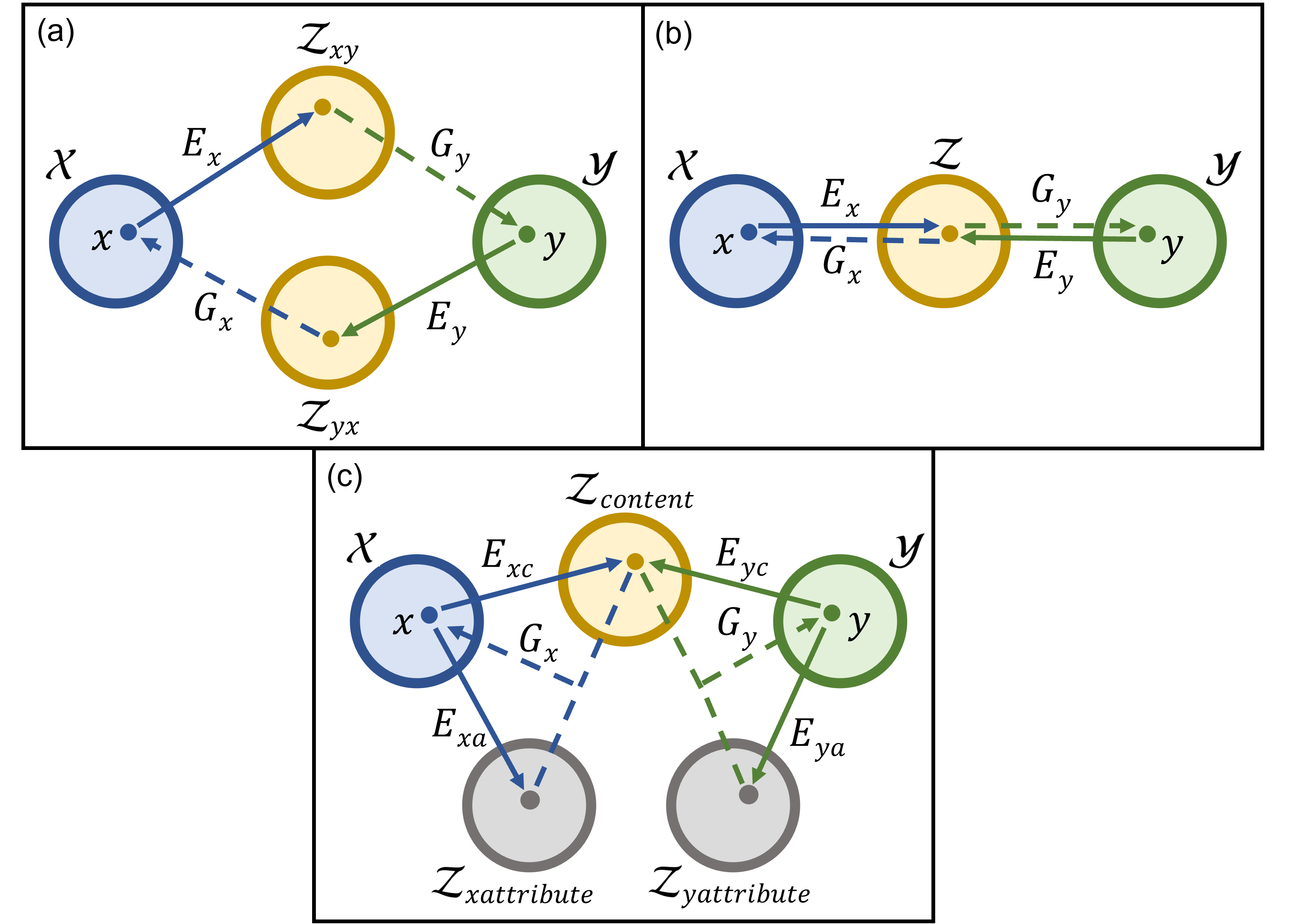}
\end{figure}

Many generative style transfer approaches use multiple generators, with a pair of generators for each pair of domains, one transferring domain A to domain B and the other transferring domain B to domain A. Examples of these approaches are shown in Figure \ref{fig:imagetoimage}. 
Multi-generator methods are particularly common in unsupervised settings, where there are no paired images available in different domains for training, making it more difficult to retain structural information in generated outputs. 

The most commonly used unsupervised method is CycleGAN \cite{Zhu2017}, a method whereby one GAN generates the stylised image, and a second GAN attempts to reconstruct the original image from the stylised image. 
A \emph{cycle consistency loss} is used to measure the similarity between original and reconstructed images:
$$ \mathcal{L}_{cyc}(G,F) = \mathbb{E}_{\mathbf{x}\sim p_{data(x)}} [||F(G(\mathbf{x}))-\mathbf{x}||_1] + \mathbb{E}_{\mathbf{y}\sim p_{data(y)}} [||G(F(\mathbf{y}))-\mathbf{y}||_1]\text{,}$$
for generators $G$,$F$. This loss is minimised to reward the accurate reconstruction of original data. This ensures that the GANs are capable of generating varied data, avoiding the issue of \emph{mode collapse}, where generated images all resemble the most common image in the modelled distribution. The CycleGAN loss is a combination of the two GAN-specific losses and the cycle consistency loss:
$$\mathcal{L}_{CycleGAN}(G,F,D_G,D_F) = \mathcal{L}_{GAN}(G,D_G) + \mathcal{L}_{GAN}(F,D_F) + \lambda\mathcal{L}_{cyc}(G,F)\text{,}$$
where $\lambda\geq0$ controls the relative importance of the different losses.

A different approach to combining two generators is taken in unsupervised image-to-image translation (UNIT) \cite{Liu2017}, with each of the domain-specific GANs sharing a latent space to learn the joint distribution of the domains instead of separately learning the marginal distributions. 
Diverse image-to-image translation via disentangled representations (DRIT) \cite{Lee2018} is a hybrid between the CycleGAN and UNIT approaches, with the latent space being decomposed into a shared part and domain-specific parts. The shared part is the \emph{content space}, which captures structure, and the domain-specific parts are the \emph{attribute spaces}, which separately capture the style of each domain. The content encoding of an image can be combined with a style encoding from a different domain to perform style transfer. This approach uses a cross-cycle consistency loss, a multi-domain extension of the cycle consistency loss, which only applies to a single pair of domains. When reconstructing the original version of an image from any style-transferred version, the generator simply switches the style embedding and leaves the content embedding unchanged. Multimodal unsupervised image-to-image translation (MUNIT) \cite{Huang2018} works similarly to DRIT, but instead of concatenating the style and content embeddings, it uses adaptive instance normalisation to combine style and content embeddings, and instead of cross-cycle consistency loss, MUNIT uses latent reconstruction losses. 


\section{Stain Normalisation}
To overcome the domain differences in histopathology datasets, researchers may use style transfer methods to normalise digital pathology images. 
Normalising training data reduces variability, which can improve convergence speed, reduce overfitting to extreme colour values, and reduce confounding effects from stain variation. Normalising during inference reduces the domain gap between in-distribution and out-of-distribution data, making it more likely that the trained model will correctly interpret the image. Stain normalisation may also have a direct clinical benefit, with the increased consistency of normalised images allowing pathologists to make diagnoses faster and with greater confidence \cite{Salvi2023}. 


\subsection{Traditional Normalisation}

Histopathological stain normalisation was already a commonly researched task before the rise of deep learning, with a variety of statistical and spectral matching techniques used \cite{Tosta2019}. Many of these approaches used colour deconvolution to separate stains, and then normalised each independently. 
Four traditional approaches have remained particularly prevalent in deep learning era research, each named after their primary author - Reinhard \cite{Reinhard2001}, Macenko \cite{Macenko2009}, Khan \cite{Khan2014}, and Vahadane \cite{Vahadane2016} normalisation. 

Reinhard normalisation \cite{Reinhard2001} was the earliest of these approaches to be developed and was the only one of the four to not have been specifically developed for use in histopathology. It is a standard statistical normalisation approach (in that it works by subtracting a mean and scaling by a variance), that works in perception-based $l\alpha\beta$ colour space, for radiance channel $l$, blue-yellow channel $\alpha$, and red-green channel $\beta$. This colour space has reduced correlation between colour channels compared to the standard red, green, and blue (RGB) space, improving the normalisation. 

Macenko normalisation \cite{Macenko2009} also converts images to a different colour space, the optical density space, a logarithmic version of the RGB space. In optical density space, stains can be linearly combined, allowing for easier separation/combination of stains. Colour deconvolution is performed using singular value decomposition to separate stain vectors from saturation values, then an estimate of the maximum intensity of each stain is then generated and used to scale all stains to have the same maximum intensity. 

Unlike the previous approaches, Khan normalisation \cite{Khan2014} is a trainable approach which requires a target domain image as well as the source image to be normalised. It uses a trainable colour deconvolution where colour-based relevance vector machine classifiers generate stain matrices for both the input and target domain image. Then, to normalise the input image, a non-linear mapping is applied to each colour channel separately based on the corresponding colour channel statistics in the target domain image. 

Vahadane normalisation \cite{Vahadane2016} is also a trainable approach which requires a target domain image. Both the source and target images undergo colour deconvolution in optical density space to separate stains using a sparse non-negative matrix factorization, which was found to be more robust to uneven stain proportions than singular value decomposition. The deconvolution approach is iteratively optimized using sparse coding and dictionary learning to generate a stain colour appearance matrix and a stain density matrix for each image, where the stain density matrix captures the locations where each stain is present, and the colour appearance matrix captures the chromatic properties of each stain. The stain density of the source is scaled using a similar approach to the intensity scaling in Macenko normalisation and then combined with the colour appearance of the target to generate a normalised source image. By retaining the stain density of the source image, this approach attempts to maintain the structure of the original image.

These traditional approaches are often beneficial to downstream AI performance when using external data \cite{Kang2021, Salehi2020, Shaban2019}. 
However, these approaches all have weaknesses and imperfections, for example, Reinhard normalisation often applies stains to background areas \cite{Khan2014, Vahadane2016, Tosta2023}. Macenko normalisation is prone to generating artifacts \cite{Khan2014} and applying stains to the wrong regions \cite{Tosta2023}. Both Macenko and Reinhard normalisations can lose structural information \cite{Vahadane2016}. Khan normalisation does not reproduce the less abundant stain(s) as well as the most abundant \cite{Vahadane2016} and can incorrectly alter the colour of dyes \cite{Sethi2016, Tosta2023}. Vahadane and Macenko normalisations both struggle to separate larger numbers of stains \cite{Kang2021}. Khan and Vahadane normalisation both use a single target image, requiring careful selection to ensure the target image is representative of the target domain as the normalised image can vary greatly based on the target image. Despite the potential issues, these are all very popular normalisation approaches which typically offer greater benefits than using non-normalised histopathology images in artificial intelligence.

\section{Generative Stain Normalisation}

In recent years, the field of deep learning has rapidly grown, with a wide array of new techniques being developed. Many recently developed stain normalisation approaches use deep learning generative AI methods. The most common model in this field is the GAN, with methods either using a single generator in a supervised setting or a pair of generators in an unsupervised setting. 



\subsection{Single Generator Normalisation Approaches}
Stain-style transfer (SST) \cite{Cho2017} is a supervised stain normalisation approach which is trained to apply stain colours to the greyscale version of an input image. This is trained using a standard GAN adversarial loss, a Euclidean distance reconstruction loss, and a Kullback–Leibler divergence feature-preserving loss. The method was evaluated for the classification of lymph node metastasis \cite{Bejnordi2017}, with slides from one institution used for training and validation, and slides from another used for testing. The proposed style transfer approach outperformed four non-GAN normalisation approaches, including Macenko and Reinhard normalisation, but did not fully overcome the domain gap, with an AUC of 0.92 compared to the target AUC of 0.98, obtained by training and testing on images from the same institution. 

This approach was iterated on by the SSIM-GAN and DSCSI-GAN methods \cite{Liang2020}, where the reconstruction losses were based on SSIM and directional
statistics-based colour similarity index (DSCSI), respectively. These reconstruction losses are more focused than the Euclidean loss, which disincentivises any change between the input and generated image. These models had slightly improved performance over SST but were still far from completely overcoming the domain gap, with SSIM-GAN and DCSCI-GAN achieving AUCs of 0.90 and 0.91, compared to a target of 0.97. 

A similar approach \cite{BenTaieb2017} used a morphological reconstruction loss based on the colour gradient field. 
In combination with a GAN adversarial loss and a cross-entropy classification loss, this approach was beneficial in multiple tasks (three classifications, one segmentation, on a total of three datasets), including outperforming Khan normalisation. Multiple sources of stain variability were investigated, with one test set digitised using a different scanner and one sourced from different pathology centres. However, this approach also did not fully bridge the domain gaps. 


Contrastive Learning for Unpaired Image-to-Image Translation (CUT) \cite{Park2020} is an approach which was not developed for stain normalisation but which has been applied to the task in subsequent research \cite{Altini2023, Zingman2023}. A \emph{patchwise contrastive loss} is used to make patch embeddings similar for patches that are drawn from the same location in the real and generated image, and dissimilar for patches drawn from different locations. The default version of the model uses an adversarial loss and an identity loss as well as the patchwise contrastive loss, with the faster version, FastCUT, not including the identity loss. The authors described FastCUT as a one-sided version of CycleGAN, allowing for much faster training with much lower memory requirements due to only using a single GAN rather than the two GANs employed in CycleGAN. 

Stain-to-Stain Translation (STST) \cite{Salehi2020} is an approach which builds on pix2pix \cite{Isola2017}, a supervised conditional GAN (cGAN) approach with an L1 reconstruction loss. 
The model was used to normalise data from one scanner to appear as though it was scanned using a different scanner, and was found to be beneficial to a greater extent than Reinhard, Macenko, Khan, and Vahadane normalisations using 10 metrics including SSIM and mean squared error, while running faster than any of the other approaches. However, the effects of this approach on any downstream AI task were not analysed. 

Colour adaptive GAN (CAGAN) \cite{Cong2022} is another greyscale-supervised pix2pix-based approach which aims to make stain normalisation more consistent. 
The proposed consistency approach uses two decoders (within the generator) with different structures to generate two stain-normalised versions of the same input image. Perturbations (augmentations) are applied to input images and to feature-space representations separately for each decoder, and the model is trained in a self-supervised fashion to overcome these perturbations, using a mean distance consistency loss between the two versions of the generated image. These generated images are also compared to the input images using a  content loss between the feature representations for the target domain, and a histogram loss based on the Hellinger distance for the source domain. These losses are combined with an adversarial loss and an L1 supervised loss between real and generated images. This complex combination of losses was found to be beneficial, with CAGAN reported to outperform traditional normalisation, STST, a non-generative network-based approach \cite{Tellez2019}, and the CycleGAN-based StainGAN for breast cancer classification 
and mutation status classification using glioma data. 
It is worth noting that all of these approaches provided large improvements over having no stain normalisation for most metrics. For five out of six metrics, CAGAN for stain normalisation also outperformed a stain augmentation approach which was previously reported to be better than a range of stain normalisation across various downstream tasks \cite{Tellez2019}. 

\subsection{Multi-Generator Normalisation Approaches}
\subsubsection*{Standard CycleGAN}
\begin{figure}[htbp]
\centering
\caption{CycleGAN stain normalisation applied to breast cancer tissue, with a different scanner used in the original and target domains. Figure adapted from \cite{Breen2022}.}
\label{fig:cyclegan}
\includegraphics[trim={0cm 0cm 0cm 0cm},width=\textwidth]{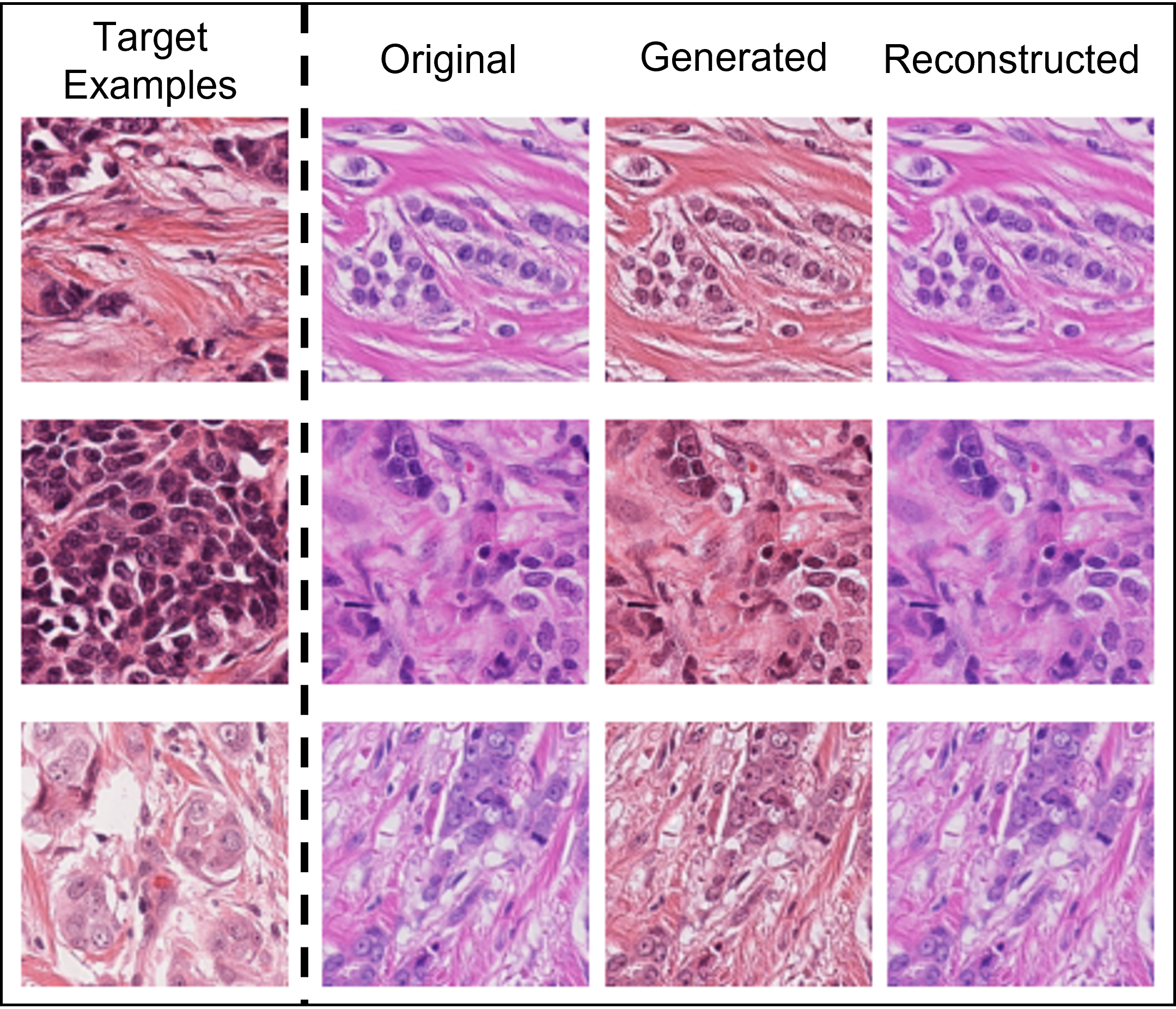}
\end{figure}

The most commonly used multi-generator approach in histopathology normalisation is the CycleGAN \cite{Zhu2017}. In our previous research \cite{Breen2022}, we evaluated normalisation approaches in the context of mitosis detection \cite{Aubreville2023}, and found that CycleGAN was only beneficial with one of our two baseline methods, with this being a marginal benefit. This indicates that CycleGAN typically generated images which were too artificial to be interpreted more accurately by the trained detection models. Despite this, CycleGAN-generated images visually appeared to be high quality, as shown in Figure \ref{fig:cyclegan}. Macenko normalisation was not beneficial for either baseline method. 

CycleGAN performance has also been found to be inconsistent  in other studies. It was found to be less beneficial than Vahadane normalisation for lymph node cancer classifications in breast tissue and colon tissue in one study including two different classifiers \cite{Stacke2020}. A different study found CycleGAN to be more beneficial than a non-generative normalisation approach \cite{Bejnordi2015} in the classification of cancer in prostate biopsies with two independent test sets \cite{Swiderska2020}. It has also been found to be beneficial compared to Macenko and Vahadane normalisations in ovarian cancer subtype classification when using external test data \cite{Shin2021}.
In each of these studies, CycleGAN gave an improvement over the baseline with no normalisation, though in our study it only provided a benefit for one of the two baseline approaches \cite{Breen2022}. 

While CycleGAN generates visually pleasing images, it has some limitations. In the context of renal glomeruli segmentation with different stains, it was found that CycleGAN-generated images belong to distinct, variable distributions, and that this has profound effects on the downstream task. It was also shown how CycleGAN can ``hallucinate", inserting new objects which do not fit the original structure \cite{Debel2021}, though this can be somewhat mitigated through the inclusion of normalisation layers in the network \cite{Vasiljevic2022}. 
CycleGAN is computationally taxing due to requiring two GANs, which is especially problematic when using data from several domains, where a different CycleGAN model needs to be trained for each pair of domains. 


\subsubsection*{Non-Standard CycleGAN}


CycleGAN can be adjusted in various ways to address the limitations, with one study finding that ``changing the generator model to a smaller U-net-like architecture, adding an identity loss term, increasing the batch size and the learning all led to improved training stability and performance" \cite{Debel2019}, which was assessed using the structural similarity index (SSIM). The identity loss was an L1 distance between the input and output of each generator to discourage large changes in the generated images, and the weight of this loss was decreased during training until it reached zero, meaning the identity loss was only used to stabilise training and avoid local optima. This adjusted CycleGAN was found to be beneficial in the context of segmenting IHC-stained renal tissue from multiple centres. 

The same authors iterated on their previous approach to generate Residual CycleGAN \cite{Debel2021} which added skip-connections from the original image to the output of the generator, focusing model training on adjusting colours rather than on the entire regeneration of images. This was also beneficial, with the authors reporting that Residual CycleGAN gave better performance than the previous approach, as well as Reinhard, Macenko, and Vahadane normalisations, StainGAN, and a lookup table normalisation approach \cite{Bejnordi2015}, in the context of colon tissue segmentation. 
It was also found that Residual CycleGAN had a reduced frequency of hallucination compared to the standard CycleGAN or StainGAN.


StainGAN \cite{Shaban2019} is an adjusted CycleGAN which uses a target style distribution rather than an individual target style image. 
This was found to improve breast cancer classification performance above that which is obtained by any of Reinhard, Macenko, Khan, or Vahadane normalisation, as well as improving generated image quality according to four similarity metrics. A transitive adversarial network (TAN) \cite{Cai2019} has been proposed as a faster alternative to this approach, where instead of a single target style image, the discriminator is provided with a new, randomly sampled style image each time it is invoked. This was found to improve SSIM over StainGAN but was not evaluated for the downstream segmentation task. 

The information learned in StainGAN can be leveraged much more efficiently using a teacher-student learning approach to distil knowledge from the CycleGAN-based StainGAN to a convolutional neural network, StainNet \cite{Kang2021}. For this approach, the StainGAN output is treated as the ground truth for the input image, allowing StainNet to be trained in a fully supervised manner using an L1 loss between the StainGAN output and the StainNet output. StainNet was found to give similar SSIM scores and downstream classification scores to StainGAN for tasks in histopathology and cytopathology. StainNet can normalise a whole slide image (WSI) in 40 seconds with much better structural similarity than StainGAN, which is prone to generating differing colours in neighbouring patches leading to an artificial checkerboard effect. 


SegCN-Net \cite{Mahapatra2020} is another CycleGAN-based approach which uses self-supervised segmentation to preserve structural information. 
The model is trained with standard cycle consistency and adversarial losses, with additional segmentation losses which compare feature map similarities between real and generated images, and between real and reconstructed images, using mean squared error. This method was found to give better classification results for lymph node classification than standard normalisation approaches (Reinhard, Macenko, Vahadane), standard CycleGAN, and two previous CycleGAN adaptations \cite{Gadermayr2018, Zhou2019}. 
It also outperformed these CycleGAN adaptations for the segmentation of glands in colorectal samples 
and outperformed standard CycleGAN according to a structural similarity index.






\subsection*{Which normalisation method is best?}

A recent comparison of stain normalisation techniques was performed in the context of colorectal cancer \cite{Altini2023}. This included Reinhard, Macenko, Khan, and Vahadane normalisations, CycleGAN, CUT, FastCUT, and two other GAN-based approaches which were not originally developed for use in stain normalisation - Geometry-consistent GAN (GcGAN) \cite{Fu2019} and AI-FFPE \cite{Ozyoruk2022}. For one dataset, 
Macenko normalisation was the best approach according to SSIM and LPIPS, and FastCUT was best according to a pixel-level similarity metric (PSNR). For the other dataset, CycleGAN was the best for each of the three metrics. The generative normalisation approaches were also evaluated subjectively by pathologists, with CycleGAN found to generate the most realistic samples for both datasets. It was not clear which method was best when considering downstream classification - out of 8 accuracy measures across 4 experiments with the generative approaches, FastCUT was best three times, CycleGAN twice, GcGAN twice, and AI-FFPE once. Macenko normalisation outperformed all generative approaches for two of the eight measures, with the best generative approach being better than the best traditional approach for the others. Further, when considering mean precision, recall, and Dice scores, FastCUT was best in some evaluations and GcGAN in others, with generative approaches typically performing much better than traditional approaches by these metrics. 

Another recent study \cite{Zingman2023} compared similar methods for stain-to-stain translation, a related task which is likely to provide useful insights for stain normalisation. The authors compared UNIT, MUNIT, pix2pix, CycleGAN, StainGAN, StainNet, CUT, Unsupervised content-preserving Transformation for Optical Microscopy (UTOM) \cite{Li2021}, and Reinhard, Macenko, and Vahadane normalisations. 
The default CycleGAN performed best in terms of the FID, and pix2pix performed best in terms of SSIM. These metrics may miss important information when taken individually, and when accounting for both metrics, StainGAN and UTOM also performed well. All of the evaluated approaches outperformed Reinhard, Macenko, and Vahadane normalisations according to FID, indicating the benefit that GAN-based approaches can offer. 

Another aspect of model selection is efficiency - many applications are resource-limited, meaning that computational complexity is an important factor. In the stain-to-stain translation study \cite{Zingman2023}, StainNet and Reinhard normalisation were by far the fastest approaches in inference regardless of available hardware. GAN-based models were particularly slow when a graphics processing unit (GPU) was unavailable, but were significantly sped up by the use of a GPU. Model training time was found to depend on the number of GANs used, with multi-GAN methods taking days to train, compared to hours for the non-generative and single GAN methods. 



Overall, it is not clear which method of stain normalisation will be best in any given scenario, with different studies finding different methods to be best \cite{Altini2023, Kang2021, Salehi2020, Sethi2016, Shaban2019,Vahadane2016, Zanjani2018a, Zingman2023}. 
A range of factors are likely to influence performance, including experimental design, evaluation approach, target domain image selection, stain type (H\&E, IHC), and other image qualities (saturation, hue, etc.).  
Ideally, then, researchers should try many different stain normalisation approaches and select the one that performs best for the given task. This is unlikely to be practical for most researchers, so when selecting, two main factors can be considered to make informed decisions. First, the state-of-the-art generative approaches often, but not always, outperform traditional approaches \cite{Altini2023, Kang2021, Salehi2020, Shaban2019, Zingman2023}. 
Second, GAN-based methods have higher computational requirements than traditional normalisation approaches, with days of GPU time often needed for the training of multi-GAN methods, and these methods taking longer in inference regardless of available hardware. GAN-based methods are unlikely to be practical in situations where GPUs are not available.






\section{Augmentation and Synthesis in Histopathology}
Generative approaches in histopathology are not only used for stain normalisation but also for augmenting training data and synthesising new data. Stain augmentation is used during model training to artificially increase the heterogeneity of staining in the training dataset, increasing model robustness. This may give better performance than normalisation in some cases, for example, one study found that using a CycleGAN for augmentation provided slightly improved segmentation results over CycleGAN for normalisation using multi-centre kidney tissue images \cite{Bouteldja2022}. Synthesis is the creation of entirely new samples from the distribution of the training data, which is another approach to artificially increase the size of the training dataset. 
Despite the potential benefits of augmentation and synthesis, a review of GANs in histopathology up to March 2021 found that normalisation was the most popular approach \cite{Jose2021}. 

\section{Conclusion}
The clinical utility of artificial intelligence in histopathology is limited by the variability of digital pathology data. It is often impractical or impossible to collect large enough repositories of varied data to train models to be truly robust to the many different sources of variability, leading to models generalising poorly on external datasets. Stain normalisation methods have been widely researched as an approach to reduce variability, from traditional statistics-based approaches to modern deep learning methods using generative AI. The most common generative approach is the generative adversarial network (GAN), typically either a single GAN in a supervised setting or multiple GANs in an unsupervised setting. GAN-based stain normalisation is beneficial for many tasks, though computational complexity is much higher than for non-generative approaches. The performance of any specific stain normalisation method varies greatly in different scenarios, with it not being clear which is the best generative approach, and with non-generative approaches outperforming generative approaches in some cases. 

\begin{acknowledgement}
JB is supported by the UKRI Engineering and Physical Sciences Research Council (EPSRC) [EP/S024336/1]. For the purpose of open access, the author has applied a Creative Commons Attribution (CC BY) licence to any Author Accepted Manuscript version arising from this submission. This is a preprint of the following chapter: Breen, J., Zucker, K., Allen, K., Ravikumar, N., Orsi, N.M., Generative Adversarial Networks for Stain Normalisation in Histopathology, published in Applications of Generative AI, Springer, edited by Lyu, Z., 2024, Springer reproduced with permission of Springer, Cham. The
final authenticated version is available online at: \url{http://dx.doi.org/10.1007/978-3-031-46238-2\_11}
\end{acknowledgement}
\addcontentsline{toc}{section}{Appendix}
%
%

\bibliographystyle{spmpsci.bst}
\bibliography{main.bib}
\end{document}